\documentclass[9pt, conference,a4paper]{IEEEtran}
\makeatletter
\long\def\@makecaption#1#2{\ifx\@captype\@IEEEtablestring%
\footnotesize\begin{center}{\normalfont\footnotesize #1}\\
{\normalfont\footnotesize\scshape #2}\end{center}%
\@IEEEtablecaptionsepspace
\else
\@IEEEfigurecaptionsepspace
\setbox\@tempboxa\hbox{\normalfont\footnotesize {#1.}~~ #2}%
\ifdim \wd\@tempboxa >\hsize%
\setbox\@tempboxa\hbox{\normalfont\footnotesize {#1.}~~ }%
\parbox[t]{\hsize}{\normalfont\footnotesize \noindent\unhbox\@tempboxa#2}%
\else
\hbox to\hsize{\normalfont\footnotesize\hfil\box\@tempboxa\hfil}\fi\fi}

\usepackage[compatibility=false]{caption}
\DeclareCaptionFont{quackfont}{\fontfamily{ptm}\fontsize{7.5pt}{9pt}\selectfont}
\usepackage[labelformat=simple,font=quackfont]{subcaption}

\makeatother
\usepackage{lipsum}
\makeatletter
\let\MYcaption\@makecaption
\usepackage{cite}
\usepackage{graphicx}
\usepackage{subcaption}
\usepackage{float}
\usepackage{refstyle}
\usepackage{multicol, blindtext}
\usepackage{amsmath,amssymb,amsfonts}
\usepackage{adjustbox}
\newcommand{\norm}[1]{\left\lVert#1\right\rVert}

\usepackage[font=scriptsize]{caption}
\usepackage[font=footnotesize]{caption}
\usepackage{stfloats}
\usepackage{lmodern,bm}                
\usepackage[T1]{sansmath}
\usepackage{xcolor}
\usepackage{algorithm, algorithmic}
\usepackage{enumitem}
\usepackage{dutchcal}

\begin{document}

\title{Learning-Based Symbol Level Precoding: A Memory-Efficient Unsupervised Learning Approach}

\author{\IEEEauthorblockN{Abdullahi Mohammad\IEEEauthorrefmark{1}, Christos Masouros\IEEEauthorrefmark{1} and Yiannis Andreopoulos\IEEEauthorrefmark{1}}\\
\IEEEauthorblockA{\IEEEauthorrefmark{1}Department of Electronic and Electrical Engineering, University College London, WC1E 7JE UK}\\[-2.0ex]
e-mail: (abdullahi.mohammad.16; c.masouros; i.andreopoulos)@ucl.ac.uk
}

\maketitle

\begin{abstract}
Symbol level precoding (SLP) has been proven to be an effective means of managing the interference in a multiuser downlink transmission and also enhancing the received signal power. This paper proposes an unsupervised-learning based SLP that applies to quantized deep neural networks (DNNs). Rather than simply training a DNN in a supervised mode, our proposal unfolds a power minimization SLP formulation in an imperfect channel scenario using the interior point method (IPM) proximal \textit{`log'} barrier function. We use binary and ternary quantizations to compress the DNN’s weight values. The results show significant memory savings for our proposals compared to the existing full-precision SLP-DNet with significant model compression of $\sim21\times$ and $\sim13\times$ for both binary DNN-based SLP (RSLP-BDNet) and ternary DNN-based SLP (RSLP-TDNets), respectively.
\end{abstract}

\IEEEpeerreviewmaketitle

\section{Introduction}
\IEEEPARstart{M}{ultiple}-input-multiple-output (MIMO) is one of the essential techniques for fifth-generation (5G) wireless communication and has recently attracted a myriad of research. Conventional block-level precoding (BLP) methods that exploit the spatial multiplexing of the multi-user MIMO system, is employed at the base station (BS) to mitigate the multi-user interference (MUI) have proven to be computationally efficient than the optimal dirty paper coding (DPC) but suffer performance deterioration \cite{windpassinger2004precoding}. The method for classifying instantaneous interference into constructive and destructive was first investigated in \cite{masouros2007novel}. The suboptimal precoding methods that exploit constructive interference (CI) were first introduced in \cite{masouros2010correlation}.\par 
The optimization-based precoding methods are intriguing because of their propensity to deliver various performance targets. The first optimization-based CI precoding was proposed in the context of vector perturbation strategy in \cite{masouros2014vector}. Additional performance is achieved by applying the precoding coefficients on a symbol-by-symbol basis termed symbol level precoding (SLP) that exploits the multiuser interference via CI with the known channel state information (CSI) and converts it into beneficial power at the receiver. Such precoding strategies have been extensively studied over the last five years. \cite{masouros2015exploiting,amadori2016constant,spano2017symbol,masouros2018harvesting}. More recently, a closed-form optimal precoding design via CI exploitation in the MISO downlink for optimization with both strict and relaxed phase rotations was proposed \cite{li2018interference}. Running  CI-based precoding methods online on a symbol-by-symbol basis can be computationally taxing despite the outstanding performance they offered. \par
With relatively low inference complexity, deep learning (DL)-based precoding designs have recently been proposed for MU-MIMO downlink transmission.  \cite{alkhateeb2018deep,xia2019deep,de2018robust,huang2019fast}. However, learning-based strategies for wireless physical layer designs use DL model as a function approximator in a supervised learning mode, which requires labeled training data. This labeled training data is obtained from the analytical solution of the optimization problem, whose accuracy is bounded by the optimization algorithm.\cite{mohammad2021unsupervised}.\par
The DL model contains millions of trained parameters, which are often stored in a 32-bit floating-point (FP32) numerical format. However, this renders the trained DL model computationally inefficient during inference and challenging to deploy to the edged devices due to resource limitations (memory and power constraints). Scalable deep neural network (DNN) models, whose weights are expressed in lower numerical precision, have been recently attracted a lot of attention \cite{rastegari2016xnor,hubara2017quantized,he2017channel}. While this idea is not new in computer vision, it has not been adequately explored within the wireless communications domain. A scalable DDN-based MIMO receiver design, where the insignificant neurons were systematically attenuated or removed via monotonically decreasing functions to reduce the network's size, was first introduced in the work of\cite{mohammad2020complexity,mohammad2020accelerated}. However, in this work, we propose an unsupervised, low precision DNN-based SLP framework, where DNN weights are constrained to binary values based on the initial work on scalable learning-based SLP designs\cite{mohammad2021memory}.
\begin{figure*}[!tbh]
\hspace{5mm}%
\subfloat[Constructive interference constellation\label{fig:CI_CONST}]{%
  \includegraphics[width=2.5in,height=1.8in]{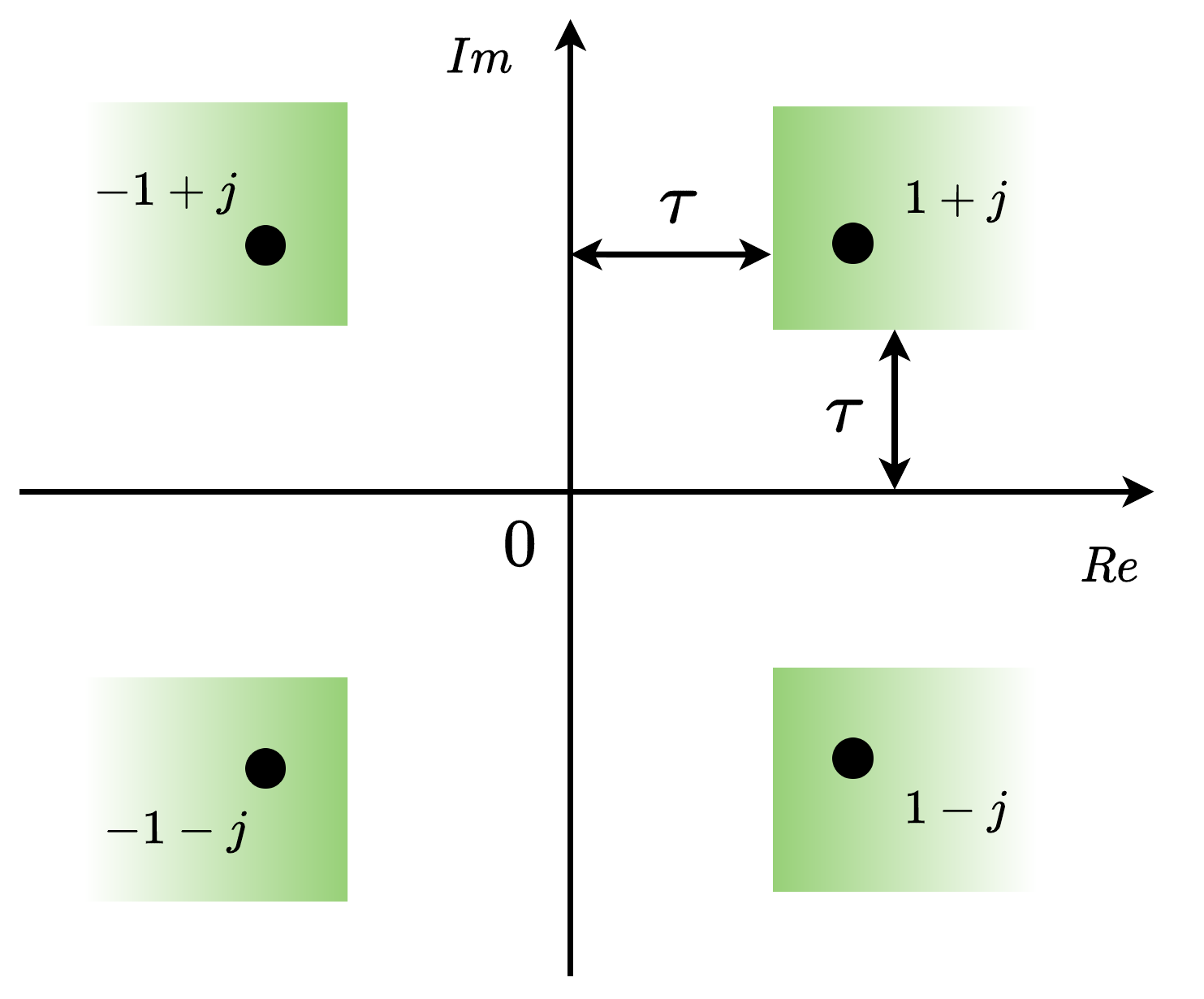}%
}
\hspace{28mm}%
\subfloat[Generic geometrical optimization regions for interference exploitation\label{fig:CI_GEOMETRY}]{%
  \includegraphics[width=2.5in,height=1.8in]{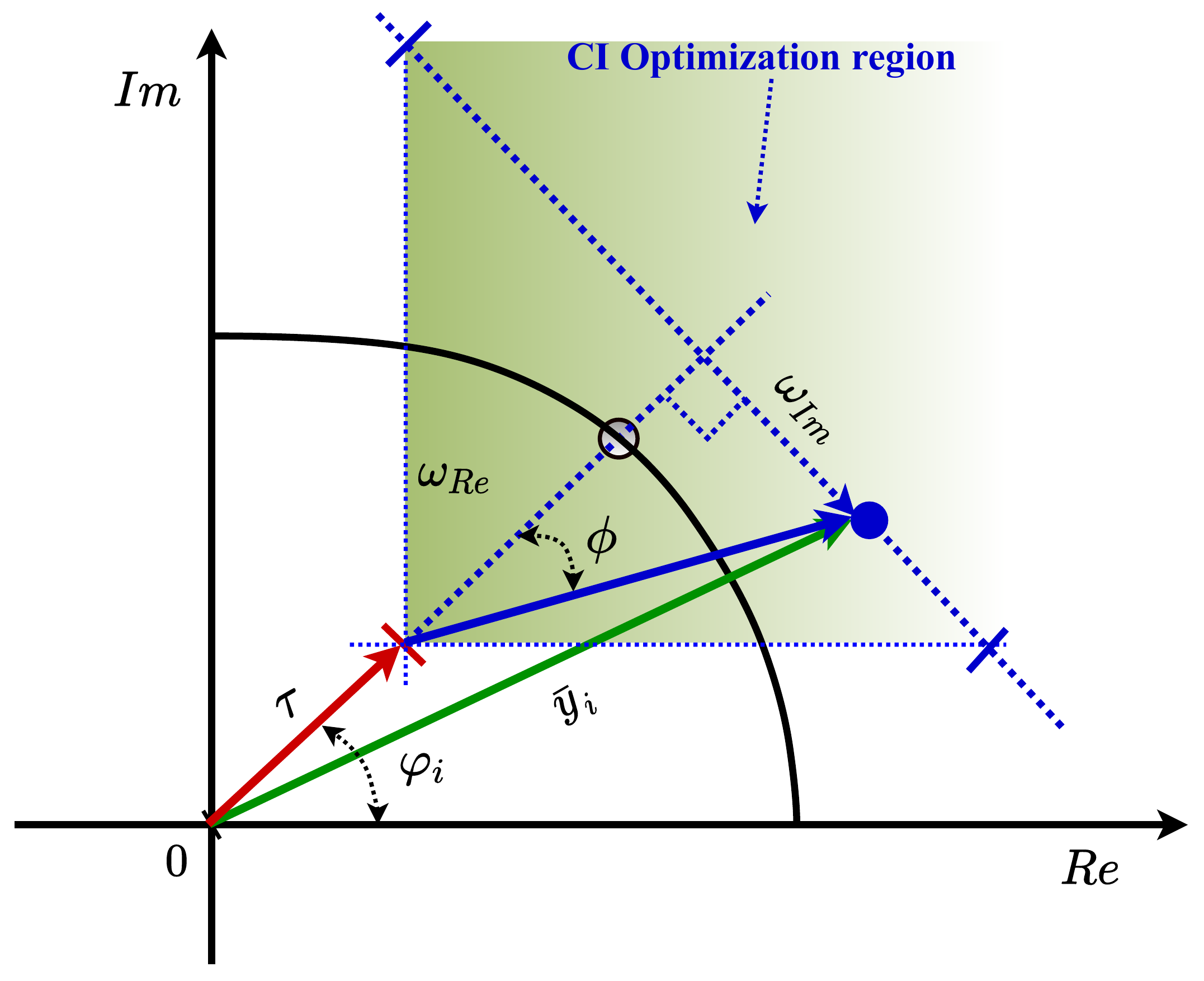}%
}
\caption{Graphical representation of Interference
Exploitation for Precoding design in QPSK \cite{masouros2015exploiting}}
\label{fig:CI_GRAHICAL_REP}
\end{figure*}

\section{System Model and Symbol Level Precoding}\label{section2}
\subsection{System Model} 
Consider single cell MU-MISO downlink transmission scenario where $K$ single-antenna users are served by an by $M$ BS antennas. Assuming a flat-fading Rayleigh channel $\mathbf{h}_{i}\in \mathbb{C}^{{N}_{t}\times1}$, the received signal at the $i$-th user is expressed as \begin{equation}\label{received_signal}
y_{i}=\mathbf{h}^{T}_{i}\sum\limits ^{K}_{k=1}\mathbf{w}_{k}{d}_{i}+n_{i},
\end{equation}
where $\mathbf{h}_{i}$, $\mathbf{w}_{i}$, ${d}_{i}$ and ${n}_{i}$ represent the channel vector, precoding vector, data symbol, phase rotation and additive white Gaussian noise for the \textit{i}-th user.\par
Instantaneous interference is categorized into constructive and destructive \cite{li2020tutorial}. As an illustration, Fig. \ref{fig:CI_CONST} shows the QPSK constellations diagram, where the CI area is indicated by the green region with respect to the minimum distance ($\tau$) from the decision boundaries, allowing the interfering signals to be added constructively with the symbol of interest via precoding vectors. The generic geometrical representation of the CI in Fig. \ref{fig:CI_GEOMETRY} shows that if the maximum angle shift ($\phi=0$) in the CI region is zero, the interfering signals overlap completely on the symbol of interest. Hence, the problem becomes a strict phase angle optimization. However, the strict phase formulation is not appealing because it leads to an additional transmission power compared to the corresponding relaxed counterpart \cite{li2020tutorial}. For simplicity, we define the following variables,
$\Hat{\mathbf{h}}_{i}=\mathbf{h}_{i}\sum_{k=1}^{K}e^{j(\phi_{k}-\phi_{i})} \in{\mathbb{C}^{M\times1}}$, ${\mathbf{w}}=\sum_{k=1}^{K}\mathbf{w}_{k} \in{\mathbb{C}^{M\times1}}$, $\Hat{\mathbf{h}}_{Ri}=\Re\{\Hat{\mathbf{h}}_{i}\}$, $\Hat{\mathbf{h}}_{Ii}=\Im\{\Hat{\mathbf{h}}_{i}\}$, $\mathbf{w}_{R}=\Re\{\mathbf{w}\}$ and $\mathbf{w}_{I}=\Im\{\mathbf{w}\}$. Similarly, we also let $\boldsymbol{\Psi} =\begin{bmatrix}
\Hat{\mathbf{h}}_{Ri}\ \Hat{\mathbf{h}}_{Ii}\
\end{bmatrix}^{T}$, $\mathbf{w}_{1} = [
\mathbf{w}_{R}\ \ -\mathbf{w}_{I}]^{T}$, where $
\boldsymbol{\Theta}=\begin{bmatrix}
\mathbf{O}_{M} & -\mathbf{I}_{M}\\
\mathbf{I}_{M} & \mathbf{O}_{M}
\end{bmatrix}\ \in \mathbb{R}^{2M \times 2M}$.
\subsection{Conventional Robust Precoding}
In practice, the exact channel state information (CSI) is often unknown; only the estimate is
\begin{equation}\label{uncert_channel}
   \hat{\mathbf{h}}_{i}={\mathbf{h}}_{i}+{\mathbf{e}}_{i}\ \forall{k},
\end{equation}
where ${\mathbf{h}}_{i}$ is the known CSI estimates at the BS and ${\mathbf{e}}_{i}$ denotes the channel error. Given this, the robust traditional recording for the downlink MU-MISO power minimization optimization is\cite{zheng2008robust}
\begin{equation}\label{conv_robust2}
    \begin{aligned}
    & \underset{\left\{\bar{\mathbf{W}}_{i} \succeq 0,\ {d}_{i} \geq 0\right\}}{\text{min}}
    & & \sum ^{K}_{i=1}\text{trace}(\bar{\mathbf{W}}_{i}) \\
    & \text{s.t.} 
    & & \begin{bmatrix}
\mathbf{\hat{\mathbf{h}}}^{*}_{i}\boldsymbol{T }_{i}\hat{\mathbf{h}}^{T}_{i} -\gamma _{i} n_{0} -{d}_{i} \delta ^{2}_{i} & \mathbf{\hat{\mathbf{h}}}^{*}_{i}\boldsymbol{T }_{i}\\
\boldsymbol{T }_{i}\hat{\mathbf{h}}_{i}^{T} & \boldsymbol{T }_{i} +\delta ^{2}_{i}\mathbf{I}
\end{bmatrix} \succeq 0 \ \forall k
\end{aligned}
\end{equation}
where $\boldsymbol{T}_{i}\overset{\Delta}{=}\bar{\mathbf{W}}_{i}-\Gamma_{i}\sum_{k=1,k\neq i}^{K} \bar{\mathbf{W}}_{k} \ \forall{k}$ and $\bar{\mathbf{W}}_{i}=\mathbf{w}_{i}\mathbf{w}_{i}^{\dagger}$.
\subsection{Robust SLP optimization-Based Power Minimization}
The multi-cast CI formulation of the power minimization problem for the worst-case scenario is given by \cite{zheng2008robust}
\begin{equation} \label{robust_multi}
    \begin{aligned}
    & \underset{\mathbf{\{w\}}}{\text{min}}
    & & {\norm{\mathbf{w}}_{2}^2} \\
    & \text{s.t.}
    & & \Bigl|\Im\{\hat{\mathbf{h}}_{i}^{T}\mathbf{w}\}\Bigl|-\left({\Re\{\hat{\mathbf{h}}_{i}^{T}\mathbf{w}\}}-\sqrt{\Gamma_{i}n_{0}}\right)\text{tan}{\phi}\leq 0,\\
    &&& \forall \norm{\Hat{\mathbf{e}}_{i}}^{2}\leq{\delta}_{i}^{2},\ \forall{i}.
    \end{aligned}
\end{equation}
For simplicity, we drop the subscripts in  (\ref{robust_multi}) and slit the real and imaginary parts of the constraint into two separate constraints real-valued functions as follows
\begin{equation}\label{robust_constraint1}
\boldsymbol{\Psi}^{T}\mathbf{w}_{1}-\boldsymbol{\Psi}^{T}\mathbf{w}_{2}\text{tan}{\phi}+\delta\norm{\mathbf{w}_{1}-\mathbf{w}_{2}\text{tan}{\phi}}_{2} + \sqrt{\Gamma n_{0}}\text{tan} \phi \leq 0,
\end{equation}
\begin{equation}\label{robust_constraint2}
-\boldsymbol{\Psi}^{T}\mathbf{w}_{1}-\boldsymbol{\Psi}^{T}\mathbf{w}_{2}\text{tan}{\phi}+\delta\norm{\mathbf{w}_{1}+\mathbf{w}_{2}\text{tan}{\phi}}_{2}+ \sqrt{\Gamma n_{0}}\text{tan} \phi \leq 0,
\end{equation}
where $\Hat{\mathbf{e}}\overset{\Delta}{=}\begin{bmatrix}
\mathbf{e}_{R} & \mathbf{e}_{I}
\end{bmatrix}^{T}$ and $\hat{\mathbf{h}}=\mathbf{h}_R+j\mathbf{h}_I+\mathbf{e}_R+j\mathbf{e}_I$. Then (\ref{robust_constraint1}) becomes
\begin{equation} \label{robust_multi2}
    \begin{aligned}
    & \underset{\mathbf{\{w_{1},w_{2}\}}}{\text{min}}
    & & {\norm{\mathbf{w}_{1}}_{2}^2} \\
    & \text{s.t.}
    & & \text{Constraints}\ (\ref{robust_constraint1})\ \text{and}\  (\ref{robust_constraint2}), \ \forall{i}\\
    &&&  \text{where}\ \ \mathbf{w}_{1}=\boldsymbol{\Pi}\mathbf{w}_{2}.
    \end{aligned}
\end{equation}
\section{Robust Low-bit DNN-based SLP for Power minimization Problem}  
This section presents robust binary and ternary DNN-based SLP models (RSLP-BDNet and RSLP-TDNet). We begin first by formulating the full-precision DNN-based SLP counterpart (RSLP-DNet). From (\ref{robust_multi2}), we define the following: $\mathbf{Q}_{1}=\left(\boldsymbol{\Theta}-\text{tan}{\phi}\mathbf{I}\right)$ and $\mathbf{Q}_{2}=\left(\boldsymbol{\Theta}+\text{tan}{\phi}\mathbf{I}\right)$. Therefore, constraints (\ref{robust_constraint1}) and (\ref{robust_constraint2}) can be written as
\begin{equation}\label{c1}
    \boldsymbol{\Lambda}^{T}\mathbf{Q}_{1}\mathbf{w}_{2}+\delta\norm{\mathbf{Q}_{1}\mathbf{w}_{2}}_{2}+\sqrt{\Gamma n_{0}}\text{tan}{\phi}\leq0
\end{equation}
\begin{equation}\label{c2}
    \boldsymbol{\Lambda}^{T}\mathbf{Q}_{2}\mathbf{w}_{2}+\delta\norm{\mathbf{Q}_{2}\mathbf{w}_{2}}_{2}+\sqrt{\Gamma n_{0}}\text{tan}{\phi}\leq0
\end{equation}
Following this, (\ref{robust_multi2}) is thus
\begin{equation} \label{robust_multi3}
    \begin{aligned}
    & \underset{\mathbf{\{w_{2}\}}}{\text{min}}
    & & {\norm{\mathbf{w}_{2}}_{2}^2} \\
    & \text{s.t.}
    & & \text{Constraints}\ (\ref{c1})\ \text{and}\  (\ref{c2}), \ \forall{i}.
    \end{aligned}
\end{equation}
\subsection{SLP using Interior Point Method}
We begin by unfolding  (\ref{robust_multi3}) using an IPM \textit{`log'} barrier function and transform it to its equivalent unconstrained sequence of sub-problems per user\cite{mohammad2021unsupervised}
\begin{equation}\label{sub_prob}
    \begin{aligned}
    & \underset{\mathbf{w \in{\mathbb{R}^{2M\times1}}}}{\text{min}} f(\mathbf{w}_{2})+{\upsilon}{{B(\mathbf{w}_{2})}},
    \end{aligned}
\end{equation}
where ${B}(\cdot)\triangleq-\sum\ln{(\cdot)}$ is the logarithmic barrier function, $\upsilon$ is the Lagrangian multiplier for inequality constraints. The learning framework is derived by defining a proximity operator of (\ref{sub_prob}) \cite{mohammad2021unsupervised} 
\begin{equation}\label{prox_op}
   \begin{aligned}
   \text{prox}_{\gamma{\upsilon}{B}}{(\mathbf{w}_{2})}= & \underset{\mathbf{w_{2} \in{\mathbb{R}^{2M\times{1}}}}}{\text{argmin}}
    & & {\frac{1}{2}\norm{\mathbf{w}_{0}-\mathbf{w}_{2}}}_{2}^{2}+\gamma{\upsilon}{B}({\mathbf{w}_{2}}), 
    \end{aligned}
\end{equation}
where $\gamma \in \{0,+\infty\}$ is the training step-size, $\mathbf{w}_{0}$ is the initial precoding vector and $\upsilon$ is the Lagrange multiplier of the inequality constraint.\par
\subsubsection{Euclidean Constraint} It can be observed that the constraints (\ref{c1}) and (\ref{c2}) are bounded by the $\mathcal{l}_{2}$-norm of the form
\begin{equation}\label{norm_constraint}
  \mathcal{C}=\{\mathbf{z}\in\mathbb{R}^{n}\big|\norm{\mathbf{z}-\mathbf{x}}_{2}\leq\alpha\},  
\end{equation}
where $\alpha>0$ and $\mathbf{x}\in\mathbb{R}^{n}$. The `{log} barrier function is given by\cite{bertocchi2020deep}
\begin{equation}\label{prox_func_robust}
{B(\mathbf{z})} = \left \{
  \begin{aligned}
    &-\ln{\left(\alpha-\norm{\mathbf{z}-\mathbf{x}}_{2}\right)}, && \text{if}\ \norm{\mathbf{z}-\mathbf{x}}_{2}<\alpha \\
    &+\infty, && \text{otherwise}.
  \end{aligned} \right.
\end{equation}
Based on (\ref{prox_func_robust}), the barrier function for (\ref{c1}) is expressed at the bottom of this page. Similar expression can also be written for (\ref{c2}). Therefore, the effective barrier function for the two constraints is the sum of the individual barrier functions ${B}(\mathbf{w}_{2})={B_{1}(\mathbf{w}_{2})}+{B_{2}(\mathbf{w}_{2})}$.\par
\begin{figure*}[!b]
\hrule
\vspace{5mm}
\begin{equation}\label{prox_func_robust1}
    {B_{1}(\mathbf{w}_{2})} = \left \{
  \begin{aligned}
    &-\ln{\left(-\sqrt{\Gamma n_{0}}\text{tan}{\phi}-\left(\boldsymbol{\Psi}^{T}\mathbf{Q}_{1}\mathbf{w}_{2}+\delta\norm{\mathbf{Q}_{1}\mathbf{w}_{2}}_{2}\right)\right)},&& \text{if}\ \boldsymbol{\Psi}^{T}\mathbf{Q}_{1}\mathbf{w}_{2}+\delta\norm{\mathbf{Q}_{1}\mathbf{w}_{2}}_{2}<-\sqrt{\Gamma n_{0}}\text{tan}{\phi} \\
    &+\infty, && \text{otherwise}
  \end{aligned} \right.
\end{equation}
\end{figure*}
It can be seen that the upper bounds of the two constraints (\ref{c1}) and (\ref{c2}) are zeros. Therefore, combining (\ref{c1}) and (\ref{c2}), we obtain
\begin{equation}\label{c5}
    \left({\delta}^{2}-\boldsymbol{\Psi^{T}\Psi}\right)\mathbf{G}\norm{\mathbf{w}_{2}}_{2}^{2}
    +4\boldsymbol{\Psi}^{T}\mathbf{w}_{2}\text{tan}{\phi}\sqrt{\Gamma{n}_{0}}\leq2\Gamma{n}_{0}\text{tan}^{2}{\phi}
\end{equation}
where $\mathbf{G}=\mathbf{Q}_{1}^{T}\mathbf{Q}_{1}+\mathbf{Q}_{2}^{T}\mathbf{Q}_{2}$. Consequently, for each $\mathbf{w}_{2}$, the proximity operator of the barrier $\gamma\upsilon{B}$ is
\begin{equation}\label{robusr_prox_opr_final}
    \Psi(\mathbf{w}_{2},\gamma,\upsilon)=\frac{2\Gamma{n}_{0}\text{tan}^{2}{\phi}-\chi(\mathbf{w}_{2},\gamma,\upsilon)^{2}}{2\Gamma{n}_{0}\text{tan}^{2}{\phi}-\chi(\mathbf{w}_{2},\gamma,\upsilon)^{2}+2\gamma\upsilon}{\mathbf{w}_{2}}
\end{equation}
where $\chi(\mathbf{w}_{2},\gamma,\upsilon)$ is the analytical solution of the cubic equation\cite{mohammad2021unsupervised}.
The robust deep-unfolded model is derived according to the derivatives of (\ref{robusr_prox_opr_final}) with respect to $\mathbf{w}_{2}$, $\gamma$ and $\upsilon$ as follows
\begin{multline}\label{jacob_mat_robust}
\mathcal{J}_{\Psi}\mid_{(\mathbf{w}_{2})}=\frac{2\Gamma{n}_{0}\text{tan}^{2}{\phi}-\norm{\Psi(\mathbf{w}_{2},\gamma,\upsilon)}_{2}^{2}}{2\Gamma{n}_{0}\text{tan}^{2}{\phi}-\norm{\Psi(\mathbf{w}_{2},\gamma,\upsilon)}_{2}^{2}+2{\gamma}{\upsilon}}\times\\
{M(\mathbf{w}_{2},\upsilon,\gamma)},
\end{multline}
\begin{multline}\label{deriv_mu_r_robustl}
    \Delta_{\Psi}\mid_{({\upsilon})}=\frac{-2\gamma}{2\Gamma{w}_{0}\text{tan}^{2}{\phi}-\norm{\Psi(\mathbf{w}_{2},\gamma,\upsilon)}_{2}^{2}+2{\gamma}{\upsilon}}\times\\
    {M(\mathbf{w}_{2},\upsilon,\gamma)}\left(\Psi(\mathbf{w}_{2},\upsilon,\gamma)\right),
\end{multline}
\begin{multline}\label{deriv_gamma_robustl}
    \Delta_{\Psi}\mid_{({\gamma})}=\frac{-2\upsilon}{2\Gamma{n}_{0}\text{tan}^{2}{\phi}-\norm{\Psi(\mathbf{w}_{2},\gamma,\upsilon)}_{2}^{2}+2{\gamma}{\upsilon}}\times\\
    {M(\mathbf{w}_{2},\upsilon,\gamma)}\left(\Psi(\mathbf{w}_{2},\upsilon,\gamma)\right),
\end{multline}
where $M(\mathbf{w}_{2},\upsilon,\gamma)$ is as defined in \cite{mohammad2021unsupervised}.\par
We use the proximity operator of the barrier to obtain the variable update function as follows
\begin{equation}\label{beam_update_robust}
\mathbf{w}_{2}^{[r+1]}=\text{prox}_{\gamma^{[r]}\upsilon^{[r]}{B}_{\text{robust}}}\left(\mathbf{w}_{2}^{[r]}-\gamma^{[r]}\Delta{f(\mathbf{w}_{2}^{[r]},\lambda^{[r]})}\right)
\end{equation}
where $f(\mathbf{w}_{2}^{[r]},\lambda^{[r]})={\norm{\mathbf{w}_{1}}}_{2}^{2}+\lambda{\mathbf{w}_{2}}$. We define the update function $\mathbb{D}$ as 
\begin{multline}
\mathbb{D}(\mathbf{w}_{2}^{[r]},\gamma^{[r]},\upsilon^{[r]},\lambda^{[r]})=\\
\text{prox}_{\gamma^{[r]}\upsilon^{[r]}{B}}\left(\mathbf{w}_{2}^{[r]}-\gamma^{[r]}\Delta{f(\mathbf{w}_{2}^{[r]},\lambda^{[r]})}\right),
\end{multline}
and $\Delta=\frac{\partial{f(\mathbf{w}_{2}^{[r]},\lambda^{[r]})}}{\partial{\mathbf{w}_{2}^{[r]}}}$.\par
\subsubsection{Loss Function} The training loss function is the Lagrangian function of (\ref{robust_multi3}) obtained  as 
\begin{multline}\label{Lag_robust1}
\mathcal{L}(\mathbf{w}_{2},\boldsymbol{\upsilon}_{1},\boldsymbol{\upsilon}_{2}) =\frac{1}{N}\sum^{N}_{i=1}\Vert \mathbf{w}_{2}\Vert_{2}^{2} \\
+\frac{\boldsymbol{\upsilon}_{1}}{N}\sum^{N}_{i=1}\left[{\delta}^{2}\norm{\mathbf{Q}_{1}\mathbf{w}_{2}}_{2}^{2}-\left(\sqrt{\Gamma n_{0}}\text{tan}{\phi}-\boldsymbol{\Psi}^{T}\mathbf{Q}_{1}\mathbf{w}_{2}\right)^{2}\right]\\
+\frac{\boldsymbol{\upsilon}_{2}}{N}\sum^{N}_{i=1}\left[{\delta}^{2}\norm{\mathbf{Q}_{2}\mathbf{w}_{2}}_{2}^{2}-\left(\sqrt{\Gamma n_{0}}\text{tan}{\phi}-\boldsymbol{\Psi}^{T}\mathbf{Q}_{2}\mathbf{w}_{2}\right)^{2}\right]\\
+\frac{\mu}{NL}\sum^{N}_{i=1} \sum_{i=1}^{L}\Vert \boldsymbol{\Omega}_{i}\Vert_{2}^{2},
\end{multline}
where $\boldsymbol{\upsilon}_{1}$ and $\boldsymbol{\upsilon}_{2}$ are the Lagrangian multipliers of the two inequality constraints. The $\boldsymbol{\Omega}_{i}(s)$ are the trainable parameters of the \textit{i}-th layers and $\mu >0$ is the penalty parameter that controls the bias and variance of the learnable parameters. Note that are associated with the barrier term and are randomly initialized from a uniform distribution. The model is trained in an unsupervised mode to update $\boldsymbol{\upsilon}$, $\lambda$, $\gamma$ and $\mathbf{w}_{2}$ such that the loss function is minimized. By minimizing (\ref{Lag_robust1}) with respect to $\mathbf{w}_{2}$, we obtain the optimal precoder
\begin{multline}\label{Lag_robust2}
    \left(1+\left(\boldsymbol{\upsilon}_{1}\norm{\mathbf{Q}_{1}}_{2}^{2}+\boldsymbol{\upsilon}_{2}\norm{\mathbf{Q}_{2}}_{2}^{2}\right)\left({\delta}^{2}-\boldsymbol{\Psi}^{T}\boldsymbol{\Psi}\right)\right){\mathbf{w}_{2}}=\\
    -\left(\boldsymbol{\upsilon}_{1}\mathbf{Q}_{1}+\boldsymbol{\upsilon}_{2}\mathbf{Q}_{2}\right)\boldsymbol{\Psi}\sqrt{\Gamma{w}_{0}\text{tan}{\phi}}.
\end{multline}
For clarity, we let $\begin{bmatrix}\| \mathbf{Q}_{1} \|_{2}^{2} & 
 \| \mathbf{Q}_{2} \|_{2}^{2}\end{bmatrix}=\bar{\mathbf{Q}}_{\text{n}}$, $\begin{bmatrix}\mathbf{Q}_{1} & \mathbf{Q}_{2} \end{bmatrix} ={\mathbf{Q'}}$ and $\begin{bmatrix}\boldsymbol{\upsilon}_{1} & \boldsymbol{\upsilon}_{2} \end{bmatrix} =\bar{\boldsymbol{\upsilon}}$.
Hence, (\ref{Lag_robust2}) is reduced to
\begin{equation}\label{Lag_robust3}
    \left(\mathbf{I}_{2M}+\mathbf{\bar{Q}}_{\text{n}}\boldsymbol{\bar{\upsilon}}^{T}\left({\delta}^{2}-\boldsymbol{\Psi^{T}\Psi}\right)\right){\mathbf{w}_{2}}=-\boldsymbol{\Psi}\mathbf{Q'}{\boldsymbol{\bar{\upsilon}}}^{T}\sqrt{\Gamma{n}_{0}}{\text{tan}{\phi}}
\end{equation}
The optimal transmit precoder is finally obtained as
\begin{equation}\label{robust_optimal}  
    \mathbf{w}_{2}=-\boldsymbol{\Psi}\mathbf{Q'}{\boldsymbol{\bar{\upsilon}}}^{T}{\mathbf{P}}^{-1}\sqrt{\Gamma{n}_{0}}{\text{tan}{\phi}},
\end{equation}
where ${\mathbf{P}}=\left(\mathbf{I}_{2M}+\mathbf{\bar{Q}}_{\text{n}}\boldsymbol{\bar{\upsilon}}^{T}\left({\delta}^{2}\mathbf{I}_{2M}-\boldsymbol{\Psi^{T}\Psi}\right)\right)$.
\subsection{RSLP-DNet and the Generic NN Architecture}
\begin{figure*}[!t]
\centering
    \includegraphics[width=\linewidth]{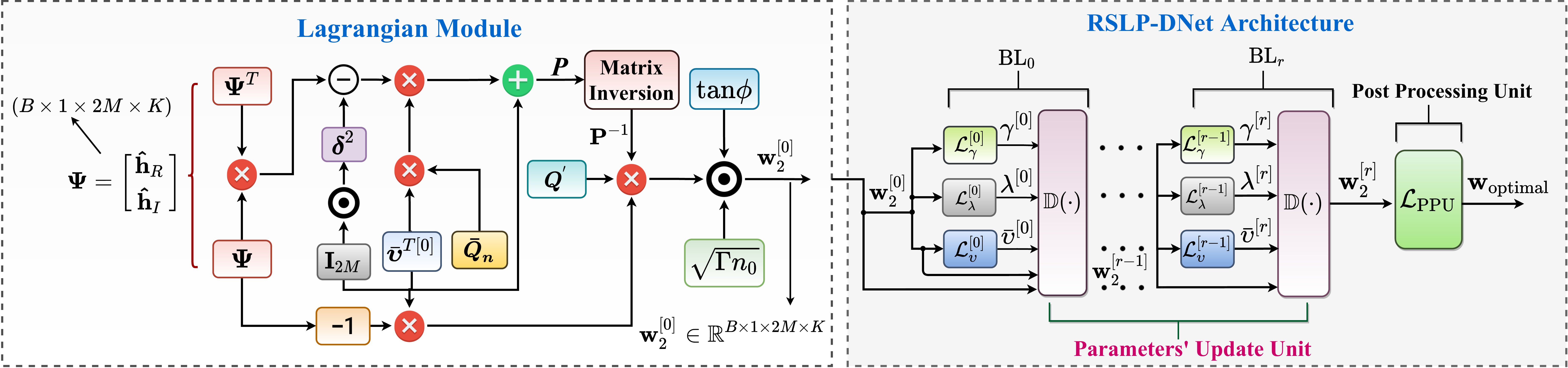}
    \caption{Complete RSLP-DNet Architecture}
    \label{fig:DNBF_Arch}
\end{figure*}

Intuitively, we can form NN cascade layers from (\ref{beam_update_robust}) as follows
\begin{equation}\label{beam_update2}
\mathbf{w}_{2}^{[l+1]}=\text{prox}_{\gamma^{[l]}{\upsilon}^{[l]}{B}}\left[\left(\mathbf{I}_{2M}-2\gamma^{[l]}\right)\mathbf{w}_{2}^{[r]}+\gamma^{[l]}\lambda^{[l]}\textbf{1}^{T}\right],
\end{equation}
where $\textbf{1}\in\mathbb{R}^{1\times2M}$ is a vector of ones. By letting $\mathbf{W}_{l}=\mathbf{I}_{2M}-2\gamma^{[l]}$, $\mathbf{b}_{l}=\gamma^{[l]}\lambda^{[l]}\textbf{1}^{T}$ and $\boldsymbol{\Xi}_{l}=\text{prox}_{\gamma^{[l]}{\upsilon}^{[l]}{B}}$, the \textit{l}-layer network $\mathcal{L}^{[l-1]}\cdots \mathcal{L}^{[0]}$ will correspond to the following
\begin{multline}\label{neural_net}
  \boldsymbol{\Xi}_{0}\left(\mathbf{W}_{0}+\mathbf{b}_{0}\right),\cdots,\boldsymbol{\Xi}_{l}\left(\mathbf{W}_{l}+\mathbf{b}_{l}\right),
\end{multline}
$\mathbf{W}_{l}$ and $\mathbf{b}_{l}$ present weight and bias parameters, respectively, and $\boldsymbol{\Xi}_{l}$ describes the nonlinear activation functions. Finally, based on this formulation, RSLP-DNet is built as shown in Fig. \ref{fig:DNBF_Arch} and its internal DNN designs are summarised in Tables \ref{tab:proximity_barrier_NN} and \ref{tab:auxiliary_NN}. 
\begin{table}[!htb]
\renewcommand{\arraystretch}{1.3}
\caption{Proximity Barrier Term DNN Design}
\label{tab:proximity_barrier_NN}
\centering
\begin{tabular}{l|l}
    \hline
    Layer  &  Parameter, $\text{kernel size}=3\times3$\\
    \hline
    \hline
    Input Layer    &  Input size $(\text{B},\ 1,\ 2M,\ K)$ \\
    \hline
    Layer 1: Convolutional    & Size $(\text{B},20,2M, K^{2})$; zero padding\\
    \hline
    Layer 2: Average Pooling    & Size $((1,\ 1),\ \text{stride}=(1,\ 1))$\\ 
    \hline
    Layer 3: Activation    & Soft-Plus \\ 
    \hline
    Layer 4: Flat
    &  Size $(\text{B}\times40\times K^{2})$ \\
    \hline
    Layer : Fully-connected  & Size$(\text{B}\times40\times K^{2},\ 1)$\\
    \hline
    Layer 5: Activation    & Soft-Plus function\\ 
    \hline
\end{tabular}
\end{table}

\begin{table}[!htb]
\renewcommand{\arraystretch}{1.3}
\caption{A PPU DNN Design}
\label{tab:auxiliary_NN}
\centering
\begin{tabular}{l|l}
    \hline
    Layer  &  Parameter, $\text{kernel size}=3\times3$\\
    \hline
    \hline
    Input Layer    &  Input size $(\text{B},\ 1,\ 2M,\ K)$ \\
    \hline
    Layer 1: Convolutional    & Size $(\text{B},\ 16,\ 2M,\ K)$,\\ & $\text{dilation} = 1$ and unit padding\\
    \hline
    Layer 2: Batch Normalization
    &  $\text{eps}=10^{-6}$, $\text{momentum}=0.1$\\
    \hline
    Layer 3: Activation   &  PReLu/k-bit function\\
    \hline
    Layer 4: Convolutional    & Size $(\text{B},\ 8,\ K,\ 2KM)$,\\ &
    $\text{dilation} = 1$ and unit padding\\
    \hline
    Layer 5: Batch Normalization
    &  $\text{eps}=10^{-6}$, $\text{momentum}=0.1$\\
    \hline
    Layer 6: Activation   & PReLu/k-bit function \\
    \hline
    Layer 7: Convolutional    & Size $(\text{B},\ 1,\ 2KM,\ 1)$,\\ & $\text{dilation} = 1$ and unit padding\\
    \hline
\end{tabular}
\end{table}
\subsection{Low-bit DNN Weights}\label{low-bit_NN}
Traditionally, DNN is designed with full-precision weights and activations. The quantization schemes have been proposed to design low-bit DNN models to address the problems of limited storage capacity and reduce hardware requirements during model deployment.
\subsubsection{1 Binary Weights:} The real-valued weights are converted to $\left(\mathbf{W}_{b} \in \{+1, -1\}^{n}\right)$. A full-precision 32-bitweight matrix is binarized such that the weights $\mathbf{W}$ are converted to their equivalent binary by the following function
\begin{equation}\label{det_binary}
    \mathbf{W}_{b}=sign(\mathbf{W})=\begin{cases}
    +1 & \text{if} \ \mathbf{W} \geq 0\\
    -1 & \text{otherwise,}
    \end{cases}
\end{equation}
A more robust binarized network ``BWN" is proposed in \cite{rastegari2016xnor} as an extension of a straightforward binary network (Binary Connect) by introducing a real scaling factor $\beta \in \mathbb{R}^{+}$ such that $\mathbf{W}\approx\beta\mathbf{W}_{b}$ by solving an optimization problem   
\begin{equation}\label{binary_optim}
   \begin{aligned}
   J(\mathbf{W}_{b},\beta)=& \underset{(\mathbf{W}_{b},\beta)}{\text{argmin}}
    & & {\norm{\mathbf{W}-\beta\mathbf{W}_{b}}_{2}^{2}} \\
    \end{aligned},
\end{equation}
and this yields
\begin{equation}\label{binary_weight}
\begin{split}
    \mathbf{W}_{b}^{*}=sign(\mathbf{W})\\
    \beta^{*}=\frac{1}{n}\norm{\mathbf{W}}_{1}
\end{split}
\end{equation}\par
\subsubsection{2 Weighted Ternary Weights:} 
\begin{figure}[!tbh]
    \centering
    \includegraphics[width=3.5in,height=1.7in]{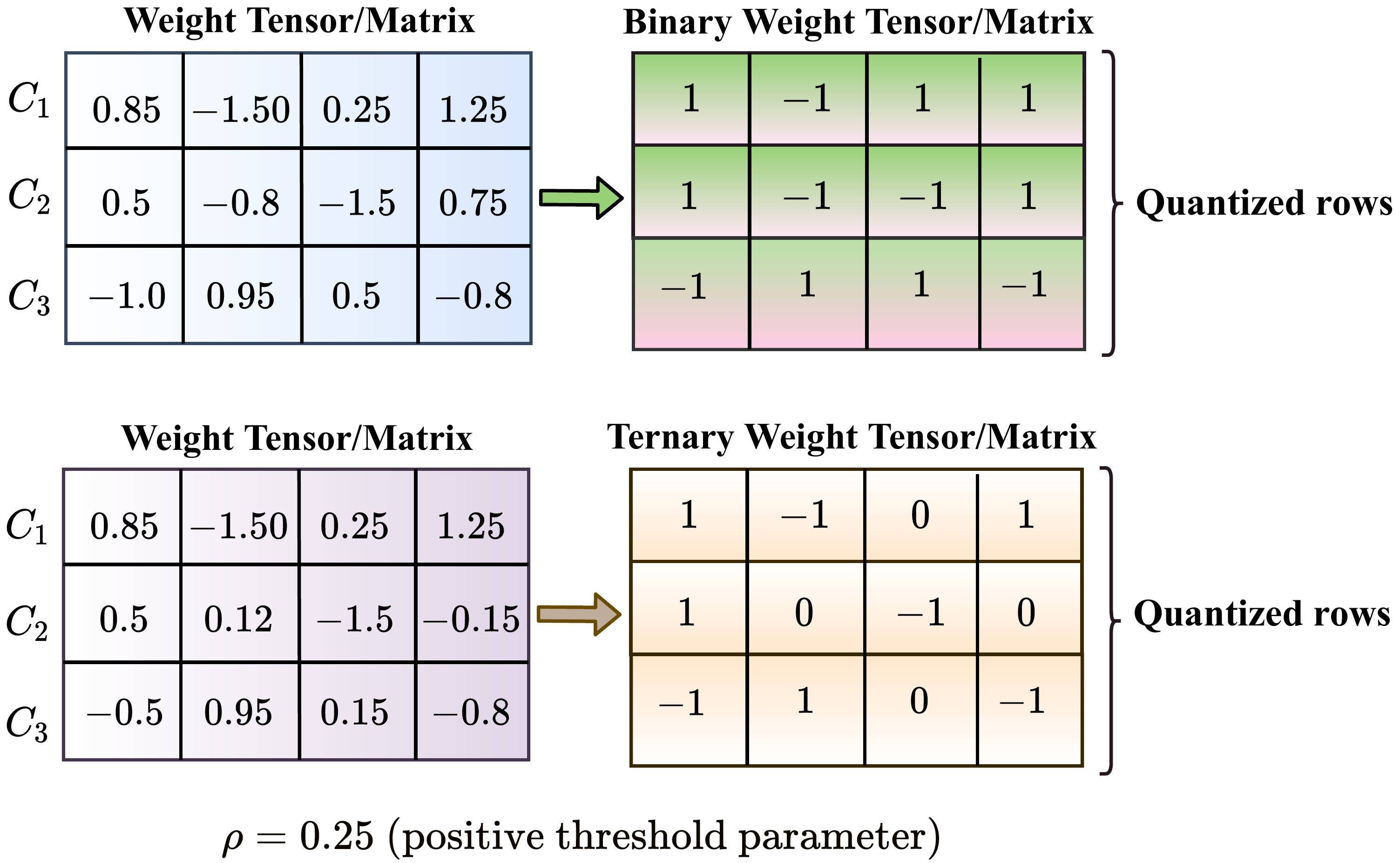}
    \caption{Binary and Ternary DNN weights}
    \label{fig:binary_ternary}
\end{figure}

A ternary weighted network (TWN) is the one in which an extra 0 state is introduced into BWN to solve the following optimization problem
\begin{equation}\label{ternary_opt}
\begin{cases}
\beta ^{*} ,\mathbf{W}^{*}_{t} = & \underset{\beta ,\ \mathbf{W}_{b}}{\text{argmin}} \ J(\mathbf{\beta ,\ W_{t}}\ ) =\| \mathbf{W} -\beta \mathbf{W}_{t} \|_{2} ^{2}\\
\text{s.t.} & \beta \geq 0,\ \mathbf{W}_{t} \in \{-1,\ 0,\ +1\}^{n},
\end{cases}
\end{equation}
and as shown in \cite{alemdar2017ternary}, solving (\ref{ternary_opt}) gives
\begin{equation}\label{ternary_weight}
\mathbf{W}^{*}_{t} =\begin{cases}
+1 & \text{, if} \ \mathbf{W}  >\rho \\
0 & \text{, if} \ |\mathbf{W} |\leq \rho \\
-1 & \text{, }\text{if} \ \mathbf{W} < -\rho,
\end{cases}
\end{equation}
where $\rho =\frac{0.7}{n}\sum\limits ^{n}_{i=1} |\mathbf{W} |$ and $\beta ^{*} =\frac{1}{\mathbf{I}_{\rho }}\sum\limits _{i\in \mathbf{I}_{\rho}} |\mathbf{W} |$,\\ $\mathbf{I}_{\rho} = \{|\mathbf{W}|>\rho\}$ is the cardinality of set $\mathbf{I}_{\rho}$.
As an illustration, Fig. \ref{fig:binary_ternary} depicts how the weight matrices are quantised based on (\ref{binary_weight}) and (\ref{ternary_weight}).
\subsection{RSLP-BDNet Training and Inference}\label{training}
The RSLP-BDNet has two central units; the parameter update unit (PUU) and the post-processing unit (PPU). The parameter unit has three core components; $\boldsymbol{\upsilon}$ (associated with the barrier term), $\lambda$ and $\gamma$ that are wired across the network (see Fig. \ref{fig:DNBF_Arch}). The barrier term is formed with one convolutional layer, an average pooling layer, a fully connected layer, and a softPlus layer to satisfy the positive inequality constraint. The PUU has \textit{r}-th blocks, each representing a layer, and is trained block-wise for \textit{l}-th iterations. Similarly, the PPU is made up three convolutional layers with batch normalisation layers in between them except for the last layer, and is trained for \textit{k}-th iterations. The number of training iterations for the PUU may not necessarily be the same as that of the PPU. We train the PUU unit for 15 iterations and the PPU 10 iterations with Adam optimizer\cite{shalev2014understanding}. We also adjust the learning rate by a factor $\vartheta \in \mathbb{R}^{+}$ for every training step to enhance the training convergence. Because the learning is done unsupervised, we use regularised Lagrangian function as a loss function. During the inference, forward pass is performed over the entire layers with the learned Lagrangian multipliers to compute the precoding vector using (\ref{robust_optimal}). The inference is performed with different SINR values to calculate the required optimal precoding matrix.  
\section{Simulation, Results and Discussion}\label{results}
\subsection{Simulation Set-up}
We assume a single cell in a downlink scenario where the BS having four antennas ($M=4$) serves $K$, single users. We generate 50000 training and 2000 test samples of channel coefficients, respectively. The transmit symbols are modulated using a QPSK modulation. The training SINR is randomly generated from uniform distribution $\Gamma_{\text{train}} \sim \mathcal{U}(\Gamma_\text{low},\,\Gamma_\text{high})$ to allow training over wide range of SINR values. parametric rectified linear unit (\textbf{PReLu}) activation function is used in RSLP-DNet instead of the traditional \textbf{ReLu} function to mitigate the effect of dying gradient due to the saturation of neurons. After every iteration, the learning rate is reduced by a factor $\alpha=0.65$ to facilitate learning convergence. The simulation parameters are summarized in Table 1. We implement the models in Pytorch 1.7.1 and Python 3.7.8 on a computer with the following specifications: Intel(R) Core (TM) i7-6700 CPU Core, 32.0GB of RAM. 
\begin{table}[!htb]
\renewcommand{\arraystretch}{1.3}
\caption{Simulation settings}
\label{tab:experimental_parameter}
\centering
\begin{tabular}{ll}
    \hline
    Parameters  &  Values\\
    \hline
    \hline
    Training Samples    &  50000 \\
    \hline
    Batch Size (B)   &  200 \\
    \hline
    Test Samples    &  2000 \\
    \hline
    Training SINR range    &  0.0dB - 45.0dB \\
    \hline
    Test SINR range (\textit{i-th} user SINR)    &  0.0dB - 35.0dB\\ 
    \hline
    Initial Learning Rate $\eta$   &  0.001 \\
    \hline
    Learning Rate decay factor $\vartheta$  &  0.65 \\
    \hline
    Number of blocks in the PUU & $B_{r}=2$ \\
    \hline
    Training Iterations for each block \\
    of the PUU   &  15 \\
    \hline
    Training iterations for PPU &  10 \\
    \hline
\end{tabular}
\end{table}
\begin{figure}[!t]
    \centering
    \includegraphics[width=2.8in,height=2.3in]{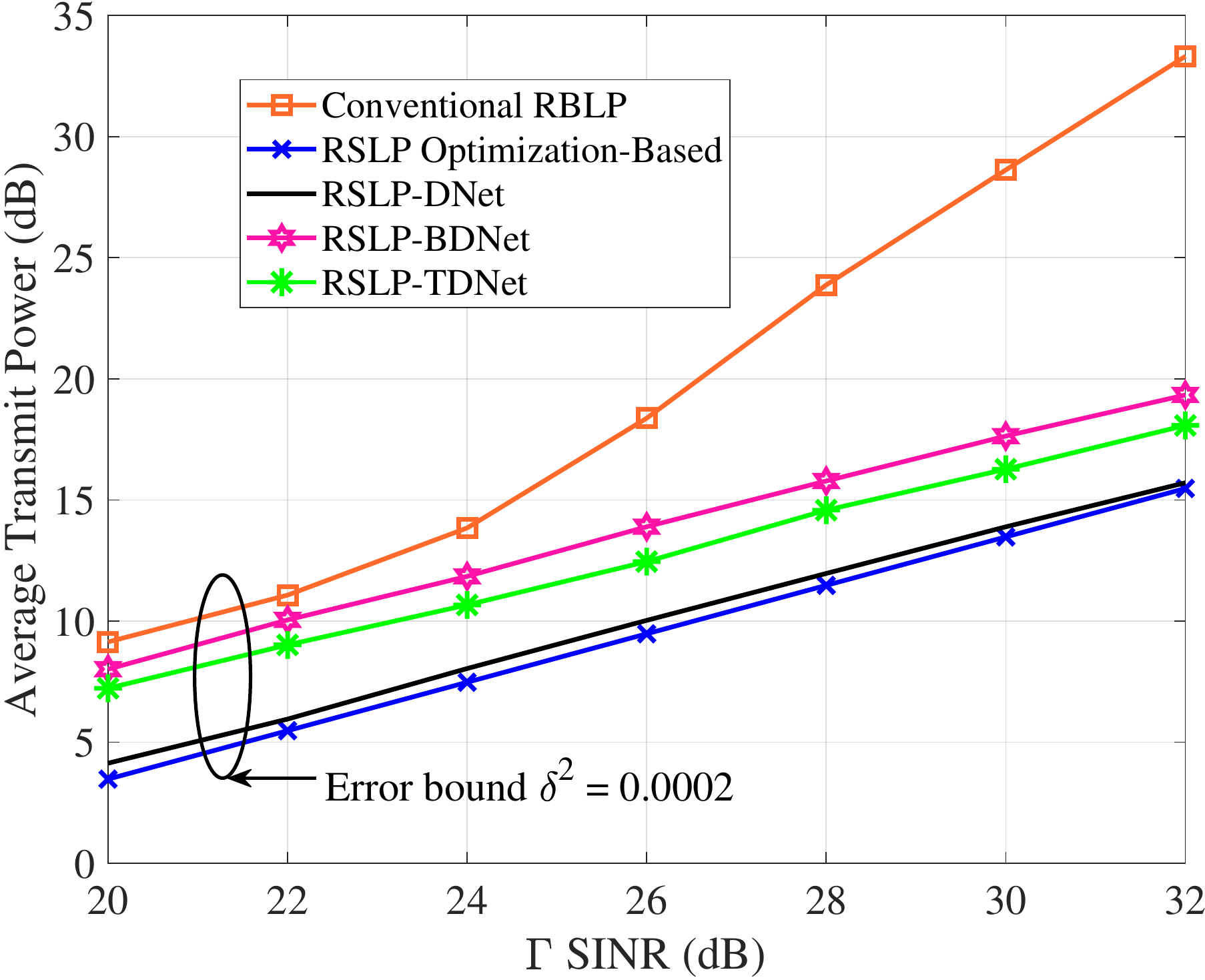}
    \caption{Transmit Power vs SINR averaged over 2000 test samples for robust conventional RBLP, RSLP optimization-based, binary and ternary DNN-based SLP solutions under $M=4$, $K=4$ and $\delta^{2}=0.0002$}
    \label{fig:POWER_vs_SINR_SQ_Robust}
\end{figure}
\begin{figure}[!t]
    \centering
    \includegraphics[width=2.8in,height=2.3in]{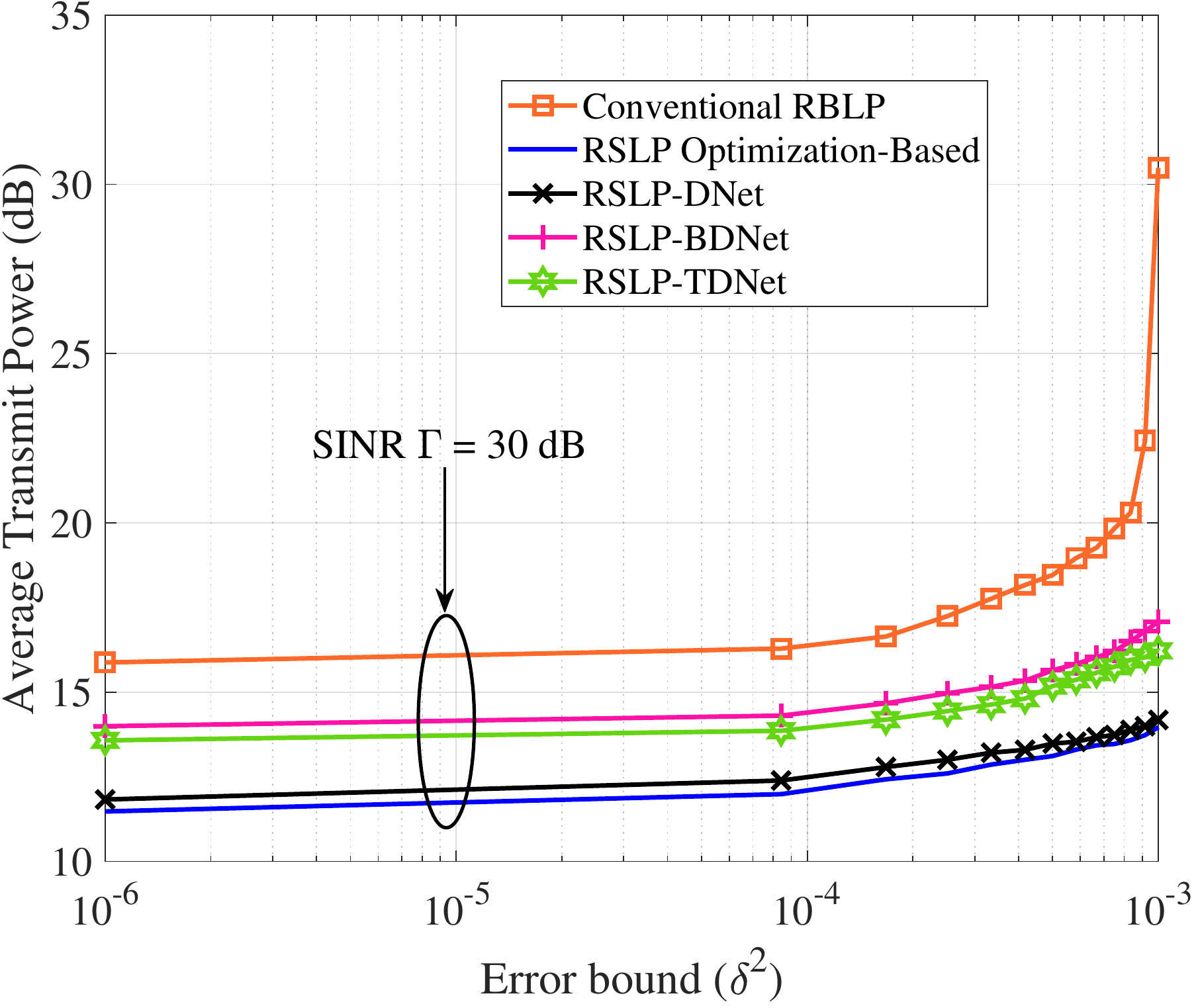}
    \caption{Transmit Power vs Error-bound for Conventional RBLP, robust RSLP optimization-based, binary and ternary DNN-based SLP solutions under $M=4$, $K=4$}
    \label{fig:POWER_vs_ERRORBOUND}
\end{figure}
\subsection{Performance Evaluation of and SLP-DNet RSLP-DNet}\label{nonrobust_performance}\label{robust_performance}
In this subsection, we consider a full-precision RSLP-DNet and its quantized counterparts (RSLP-BDNet and RSLP-TDNet). We use $4\times4$ MISO system with CSI error bounds $\delta^{2}={10}^{-4}$, and QPSK modulation scheme. We compare the average transmit power of RSLP-BDNet and RSLP-TDNet against R-SLP-DNet \cite{mohammad2021unsupervised} robust SLP optimization-based \cite{masouros2015exploiting} and conventional \cite{zheng2008robust} BLP methods. Fig. \ref{fig:POWER_vs_SINR_SQ_Robust} depicts how the average transmit power increases with the $SNR$ thresholds. The RSLP optimization-based is observed to show a significant power savings of more than 60\% compared to the conventional RBPL solution. Similarly, the proposed RSLP-BDNet and RSLP-TDNet show considerable power savings of $40\%-58\%$ against the conventional RBLP but lower lower than the RSLP optimization-based solution. \par 
Furthermore, we study the effect of the CSI error bounds on the transmit power at 30dB. Fig. \ref{fig:POWER_vs_ERRORBOUND} depicts the variation of the transmit power with increasing CSI error bounds. A significant increase in transmit power can be observed where the channel uncertainty lies within the region of CSI error bounds of $\delta^{2}=10^{-3}$. Interestingly, like the RSLP optimization-based, by exploiting the CI, the proposed methods show a descent or moderate increase in transmit power.

\begin{figure}[!t]
    \centering
    \includegraphics[width=2.8in,height=2.3in]{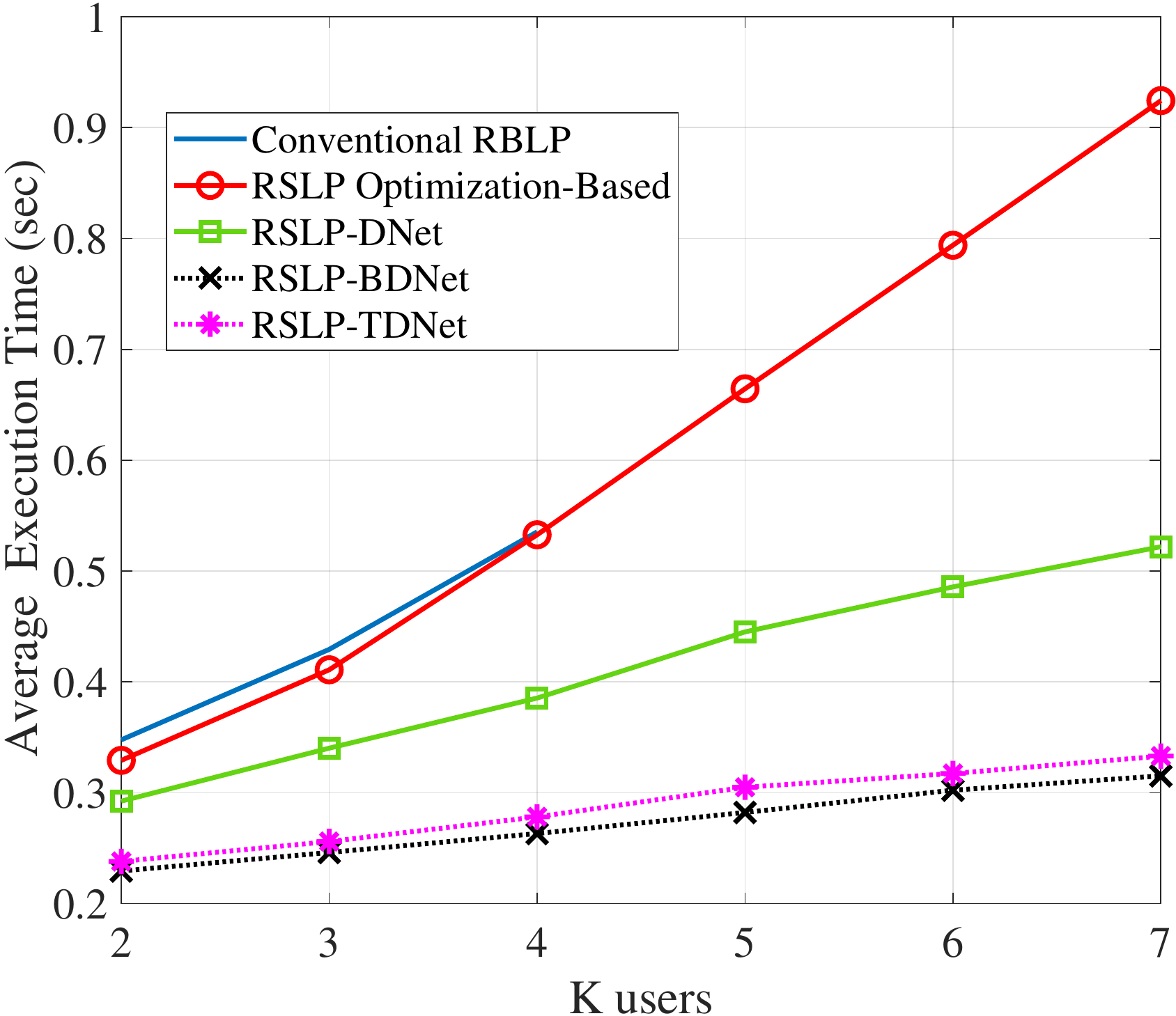}
    \caption{Comparison of average execution time per sample averaged over 2000 test samples for Nonrobust and Robust precoding schemes, i.e, conventional BLP, SLP optimization-based and SLP learning-based models under $M=4$, $K=4$}
    \label{fig:Execusion_time}
\end{figure}

\subsection{Computational Complexity Evaluation}
The computational costs of the proposed learning methods are obtained from the PUU and the feed-forward convolutions of the CNN that makes up the PPU. For the PUU, the dominant computation is in computing the proximal barrier functions, which requires computing the \textit{`log'} of the barrier and the shallow CNN structure. It can be seen that both RSLP optimization-based and the proposed schemes are feasible for all sets of $M$ BS antenna and $K$ mobile users. However, for conventional RBLP, the solution is only feasible for $M \geq K$.\par 
Fig. \ref{fig:Execusion_time}(a) shows the average execution time of the proposed learning solutions per symbol averaged over 2000 test samples. We observe that while RSLP-DNet is faster than both RSLP optimization-based and conventional RBLP, RSLP-BDNet and RSLP-TDNet offer much less average execution time by $\sim3\times$ and $\sim5\times$ compared to RSLP-DNet, respectively. This because most of the MACs operations are replaced by binary bit-wise operations. RSLP-TDNet is slightly slower than RSLP-BDNet due to the additional {`0'} binary state introduced in the former.
Accordingly, we binarized DNN could offer significant training and inference accelerations while offering good trade-off between the performance and computational complexity.

\begin{table*}[!ht]
\renewcommand{\arraystretch}{1.2}
\caption{Inference memory utilization comparison}
\label{tab:complexity}
\centering
\begin{tabular}{llllllll}
    \hline
    \hline
    Models & Weights & Activations  & Memory  & Memory\\
    & & & usage (MB) & saving & \\
    \hline
    \hline
    RSLP-DNet & $(32-\text{bit})\in\mathbb{R} $ & $(32-\text{bit})\in\mathbb{R}$ & 0.1898 & $-$\\
    \hline
    RSLP-BDNet & $\{-1,+1\}$ & $\{-1,+1\}$ & 0.0089 & $21.33\times$\\
    \hline
    RSLP-TDNet & $\{-1, 0, +1\}$ & $\{-1,+1\}$ & 0.0146 & $13\times$\\ 
    \hline
    \hline
\end{tabular}
\end{table*}
The size of the DNN is often bounded by the available memory. Therefore, it is beneficial to estimate the memory requirements of the DNN at the inference. We examine and analyze the memory utilization of full-precision RSLP-DNet and its corresponding quantized versions at inference. We adopt the approach presented in \cite{bethge2019binarydensenet} to calculate the inference memory utilization as the summation of 32-bit times the number of floating-point parameters and 1-bit times the number of binary parameters, expressed as $\frac{1}{32}W_{b}+ W_{f}$, where $W_{b}$ and $W_{f}$ are the binary and floating-point weights, respectively. Table \ref{tab:complexity} presents the summary of the inference memory requirements, where we observe tha RSLP-BDNet and RSLP-TDNet provide considerable memory savings up to $\sim21\times$ and $\sim13\times$ compared to the RSLP-DNet, respectively.
\section{Conclusion}\label{conclusion}
This paper proposes robust binary and ternary unsupervised learning-based SLP designs for downlink power minimization optimization. The real-valued NN weights are converted to binary values, allowing the operations between the inputs and weights tensors to be performed in binary operations. We use domain knowledge to design unsupervised learning architectures by unfolding the proximal interior point method barrier \textit{`log} function for a relaxed phase rotation. The performance is within the range of $89\%-95\%$ of the RSLP optimization-based solution with a substantial computational complexity reduction. Therefore, our proposals demonstrate an indispensable balance between the performance and the computational complexity involved.  
\bibliographystyle{IEEEtran}
\bibliography{ref}

\end{document}